\documentclass[fleqn,12pt]{article}
\newcommand{\AmS}{{\protect\the\textfont2
  A\kern-.1667em\lower.5ex\hbox{M}\kern-.125emS}}

\title{Reactor as a Source of Antineutrinos: \\ Thermal Fission Energy}

\author{V. Kopeikin\thanks{kopeykin@polyn.kiae.su}, L. Mikaelyan\thanks{mikaelyan@polyn.kiae.su}, 
V. Sinev\thanks{sinev@polyn.kiae.su}}
\date{Russian Research Centre "Kurchatov Institute", Moscow}

\begin{document}

% typeset front matter
\maketitle

%\tableofcontents

\begin{abstract}
Deeper insight into the features of a reactor as a source of antineutrinos is required for making further advances in studying 
the fundamental properties of the neutrino. The relationship between the thermal power of a reactor and the rate of the chain 
fission reaction in its core is analyzed.

\end{abstract}

\section*{Introduction}

Experiments aimed at studying the fundamental properties of the neutrino and at testing the standard model of electroweak 
interactions are being performed at reactors. A collaboration of researchers from the Kurchatov Institute and the Petersburg 
Nuclear Physics Institute (PNPI, Gatchina) are conducting an experiment devoted to searches for the neutrino anomalous 
magnetic moment [1]. A group of physicists from the Institute of Theoretical and Experimental Physics (ITEP, Moscow) and 
the Joint Institute for Nuclear Research (JINR, Dubna) are preparing a similar experiment at the reactor of the Kalinin atomic 
power plant [2]. The CHOOZ experiment [3], completed quite recently, set constraints on the neutrino-mixing-matrix element
$U_{e3}$. The KamLAND Collaboration, which is recording antineutrinos at a distance of a few hundred kilometers from 
reactors, is able to determine the remaining two mixing-matrix elements $U_{e1}$ and $U_{e2}$  and to test the LMA MSW 
hypothesis of solar-neutrino oscillations [4]. In addition, it should be noted that a program of neutrino studies at the reactors on 
the Taiwan island is being developed [5] and that interesting proposals concerning searches for neutrino oscillations were put 
forth in Germany [6]. (More details on the motivation of those investigations, their status, and their prospects can be found, for 
example, in the review articles cited in [7].)
 
Differing in many respects, the aforementioned experiments possess one common feature: the results obtained in these experiments 
are analyzed by an absolute method $-$ specifically, the measured counting rates for neutrino events and their spectral distributions
are contrasted against their counterparts calculated on the basis of the theory of electroweak interactions. For input data in these 
calculations, use is made of the set of features of neutrino radiation that, together with other data, form a metrological basis of the 
experimental physics of neutrinos at nuclear reactors.

The spectral density $f(E_\nu)\;({\rm cm^{-2} s^{-1} MeV^{-1}})$ of the flux of reactor electron antineutrinos 
( $\bar{{\nu}_e}$) incident on a detector is given by

\begin{equation}
f(E_\nu) = N_f {\rho_f (E_\nu)}/ 4\pi R^2,
\end{equation}
where $N_{f}$ is the number of fission events in a reactor per second, ${{\rho}_f}({E_{\nu}})$ 
$({\rm MeV^{-1} fiss.^{-1}})$ is the spectrum of reactor electron  antineutrinos that is normalized to a fission event, 
and $R$ (cm) is the distance between the reactor and the detector used.

In the fission of uranium and plutonium nuclei and in the subsequent radioactive decay of fission fragments, as well as in 
accompanying neutron reactions, energy is released, its major part being absorbed in the reactor, whereby it is converted into 
heat. Denoting by $E_f $ (MeV/fiss.) the energy absorbed in the reactor on average per fission event, we can represent 
the chain-reaction rate $N_f $ in the form

\begin{equation}
N_f  = W/E_f,
\end{equation}

The present study is devoted to exploring the quantity $E_f$ , which relates the fission-reaction rate $N_f$ (fiss./s) to the thermal 
power $W$ of a reactor. First of all, we consider this relationship for the example of a standard operating period of a reactor 
belonging to the PWR type, in which case isotopes undergoing fission include $^{235}$U, $^{239}$Pu, $^{238}$U, and 
$^{241}$Pu. The method developed here and, upon introducing some specific corrections, the results presented below can be 
used in neutrino experiments, both those that are being presently performed and those that are planned, at reactors of any other 
type.

The reactor staff determines the current value of the thermal power to a precision of about 1 to 2\%. In order to avoid increasing 
the error in determining the ratio in (2), we will try to calculate the energy $E_f$ to a higher precision.

\section{COMPONENTS OF THE ENERGY $E_f$}

The energy $E_f$ can be represented as the sum of four terms; that is,

\begin{equation}
E_f  = E_{tot}- \langle {E_\nu} \rangle - \Delta E_{\beta \gamma} + E_{nc},
\end{equation}
where  $E_{tot}$ is the total energy released in nuclear fission from the instant at which the neutron inducing this fission process 
is absorbed to the completion of the beta decays of product fragments and their transformation into beta-stable neutral atoms, 
$\langle E_{\nu} \rangle$ is the mean energy carried away by the antineutrinos that are produced in the beta decay of fission 
fragments ($\sim 6\bar{{\nu}_e}/fiss.$), $\Delta E_{\beta \gamma}$ is the energy of beta electrons and photons from 
fission fragments that did not decay at a given instant of time, and $E_{nc}$ is the energy absorbed upon neutron capture 
(without fission) in various materials of the reactor core.

That part of the total energy $E_{tot}$ which remains in the reactor and which transforms into heat forms the effective 
fission energy $E_{eff}$,

$$
E_{eff} = E_{tot} - \langle {E_\nu} \rangle - \Delta E_{\beta \gamma}. \qquad \qquad \qquad \qquad \qquad \qquad \qquad 
\eqno (3.1)
$$
The expression for $E_{eff}$ can then be represented in the form

$$
E_f = E_{eff} + E_{nc} . \qquad \qquad \qquad \qquad \qquad \qquad \qquad \qquad \qquad \eqno (3.2)
$$

The above concerns the energy released in a single nuclear-fission event, but the chain fission reaction in a reactor proceeds over 
a finite time interval. For this reason, we consider a chain fission reaction that begins at the instant $t = 0$ and proceeds at the rate
of $N_f$ = 1 fiss./s. We denote by $E(t)_{tot}$ the energy released per second at the instant $t$ reckoned from the 
commencement of the process being considered. The quantity $E(t)_{tot}$ includes all kinds of energy, with the exception of 
$E_{nc}$, which is the energy that is released in various materials upon the absorption in them of neutrons not involved in the 
fission process. We will now consider the function $f_{tot}(t)$ determining the energy released per unit time after the lapse of 
the time $t$ since a single fission event. It is obvious that

\begin{equation}
E(t)_{tot} = \int_0^tf_{tot}(t')dt', \qquad f_{tot}(t) = \frac{dE(t)_{tot}}{dt} 
\end{equation}
The energy $E(t)_{tot}$ grows with increasing fission process time $t$, tending to the limiting value $E(t)_{\infty}$,

$$
E(\infty)_{tot}= \int_0^\infty f_{tot}(x)dx \equiv E_{tot}. \quad \qquad \qquad \qquad \qquad \qquad \qquad \eqno (4.1)
$$
The above equations relate the energy release in a single fission event to the energy release per unit time in a continuous process.

Neutrino investigations are performed at reactors where use is made of uranium whose enrichment in $^{235}$U is low. As this 
isotope burns out, $^{239}$Pu and $^{241}$Pu are accumulated in the core of such a reactor. Just like $^{235}$U, these isotopes 
undergo fission induced by thermal neutrons. There is also a contribution to the total number of fission events from $^{238}$U, 
which is fissile under the effect of fast neutrons. Therefore, we have

\begin{equation}
E_f = \sum {\alpha_i E_{fi}}, \qquad  \sum \alpha_i = 1,
\end{equation}
where $\alpha_i \;(i = 5, 9, 8, 1) - $ are the contributions of the $^{235}$U, $^{239}$Pu, $^{238}$U and $^{241}$Pu isotopes to 
the total number $N_f$ of fission events at a given instant of time. Information about $\alpha_i $ values, which change in the
course of reactor operation, is provided by the reactor staff with a relative error of 5\%. The $\alpha_i $ values typical of PWR 
reactors are

\begin{equation}
\alpha_5 = 0.59, \qquad \alpha_9 = 0.29, \qquad \alpha_8 = 0.07, \qquad \alpha_1 = 0.05.
\end{equation}

It should be emphasized that the energy $E_f$ and the calculated number $W/E_f$ of fission events occurring in a reactor at 
a given instant of time are not determined exclusively by the current reactor state, which is specified by the level of the reactor 
power and by the isotopic composition of the burning nuclear fuel, but they are dependent on the prehistory of the reactor. This 
dependence is controlled by the terms $E_{nc}$ and $\Delta E_{\beta \gamma}$ which appear in (3). The quantity $E_{nc}$ 
changes along with the composition of the materials in the reactor core in the course of reactor operation. Both terms involve 
a contribution from longlived beta emitters and depend on the duration of the irradiation of the fuel.

For the isotopes undergoing fission, the energies $E_{fi}$ in (5) for the whole reactor exceed 200 MeV/fiss. Going somewhat 
ahead, we note that, for a PWR reactor, the absolute values of the terms appearing in expression (3) for the energy $E_f$ are 
in the following ratio:

\begin{equation}
E_{tot}: \langle{E_\nu}\rangle : \Delta E_{\beta \gamma}: E_{nc} \approx 200 : 9 : 0.3 : 10 \,. 
\end{equation}

\section{TOTAL $(E_{tot})$ AND EFFECTIVE $(E_{eff})$\quad \\ FISSION ENERGY }

\subsection{$Total\; Fission\; Energy\; E_{tot}$}

The energy $E_{tot}$ can be calculated by summing the mean values of various components of the energy release, such as 
the fragment kinetic energy, the energy of prompt and delayed fission gamma rays, and the neutron and beta-electron kinetic 
energies. However, much more precise results are obtained by directly applying the energy-conservation law to the fission
process; that\,is,

\begin{equation}
M(A_0,Z_0) + M_n = \sum y_A\, M(A,Z_A) + n_f\, M_n + E_{tot},
\end{equation}
where $M (A_0,Z_0)$ is the atomic mass of the isotope undergoing fission (the speed of light is set to unity, c = 1); $A_0$ and 
$Z_0$ are its mass and charge numbers, respectively; $M_n$ is the neutron mass; summation is performed over the mass numbers 
$A$ of beta-stable fission products; $M(A,Z_A)$ are the masses of these products; $y_A$ are their total yields, 
$\sum y_A = 2$; and $n_f$ is the mean total number of prompt and delayed fission neutrons (for the obvious reason, the 
notation $\nu$, which is usually used for the mean number of fission neutrons, is replaced here by $n_f$.)

Using the condition requiring that the number of nucleons be conserved in the fission process and introducing the mass excesses 
for atoms, $m(A,Z)$, we can recast relation (8) into form

\begin{equation}
E_{tot}= m(A_0,Z_0) - \sum y_A\,m(A,Z_A) - (n_f - 1)\,m_n 
\end{equation}
where\quad $m(A,Z) = M(A,Z) - A\,m_0$\quad ($m_0$ is an atomic mass unit) and\quad $m_n = M_n - m_0 = {\rm 8.0713 
\pm 0.0001}$ MeV is the neutron mass excess.

The calculated total energy $E_{tot}$ and the quantities appearing in relation (9) are given in Table 1 for all four nuclei 
undergoing fission. In computing these results, we employed data on the mass excesses for the atoms involved [8] and on 
the yields of fission fragments [9] whose mass numbers took values in the range between 66 and 172 (see Fig. 1). The data 
on the number of fission neutrons were borrowed from [10].

\begin{table}[htb]
\caption{Mass excesses and total fission energy $E_{tot}$ (in MeV/fiss.)}
\label{table}
\vspace{10pt}
\begin{tabular}{c|c|c|c|c|c}
\hline
Fissile & Mass & Mass excess & Number & $(n_f -1)m_n$ & Total \\
nucleus & excess  &  for fission & of fission & & fission \\
 & $ m(A_0,Z_0)$ & products, & neutrons,  & & energy, \\
 & & $\sum y_A\,m(A,Z_A)$ & $n_f$ & & $E_{tot}$ \\ 
\hline
$^{235}$U & 40.914 & -173.43 & 2.432 & 11.55 & 202.79 \\
& $\pm 0.002$ & $\pm 0.05$ & $\pm 0.0036$ & $\pm 0.03$ &  $\pm 0.06$ \\
\hline
$^{238}$U & 47.304 & -173.39 & 2.829 & 14.76 & 205.93 \\
& $\pm 0.002$ & $\pm 0.10$ & $\pm 0.011$ & $\pm 0.09$ &  $\pm 0.13$ \\
\hline
$^{239}$Pu & 48.584 & -173.87 & 2.875 & 15.13 & 207.32 \\
& $\pm 0.002$ & $\pm 0.07$ & $\pm 0.0060$ & $\pm 0.05$ &  $\pm 0.08$ \\
\hline
$^{241}$Pu & 52.951 & -173.72 & 2.937 & 15.63 & 211.04 \\
& $\pm 0.002$ & $\pm 0.10$ & $\pm 0.0073$ & $\pm 0.06$ &  $\pm 0.12$ \\
\hline
\end{tabular}\\[2pt]
\end{table}

For the fissile nuclei being considered, the values of $E_{tot}$ differ from one another by a few MeV, increasing in the order of 
their positions in the first column of Table 1. These distinctions are caused, above all, by an increase in the mass excess for the
atoms of the fissile isotopes and, to a lesser extent, by an increase in the number $n_f$ of fission neutrons. At the same time, 
it can be seen from Table 1 that, for the set of stable fission fragments, the total mass excess $\sum y_A\,m(A,Z_A)$ is virtually 
independent of the nucleus undergoing fission. This is because the quantity $m(A,Z_A)$ is approximately constant over the 
region of high fragment yields $y_A$, sizably increasing only for products originating from highly asymmetric fission, where the 
yields in question are relatively low (see Fig. 1). Therefore, even significant distinctions between the mass distributions of 
fragments produced in the fission of uranium and plutonium nuclei have but a slight effect on the sums $\sum y_A\,m(A,Z_A)$.

The error in the mass excess $\sum y_A m(A,Z_A)$ (see the third column in Table 1) depends on the uncertainty in the yields 
$y_A$, since the overwhelming majority of the values of\, $m(A,Z_A)$ are known to a precision not poorer than 5\,keV. In 
order to find this error, each of the yields $y_A$ was varied individually, irrespective of the others, under the assumption that it 
obeys the a Gaussian distribution. Upon each variation, there arises a new set of $y_A$ values, and we calculated the value of 
$\sum y_A m(A,Z_A)$ for this set. As a result, the total number $\sum y_A A$ of nucleons contained in fission products 
changed somewhat. On the basis of the relation

\begin{equation}
A_0 +1 = \sum y_A A + n_f, 
\end{equation}
which expresses the law of nucleon-number conservation, we calculated the corresponding number $n_f$ of neutrons. A point 
in the plane spanned by the variables $\sum y_A m(A,Z_A)$ and $n_f$ was associated with the pair of values found in this way 
for the mass excess and the number of neutrons. The results of one such computational experiment performed for $^{235}$U, 
where use was made of a Gaussian distribution characterized by a FWHM value of 0.12, are illustrated in Fig. 2 (ten thousand 
points). From Fig. 2, it can be seen that the uncertainties in the yields of fission products introduce an error of about 35 keV in 
the mass excess and that available experimental data on the yields of fission products and on the number of neutrons are quite 
consistent.

That the calculation of the total energies $E_{tot}$ on the basis of applying the energy-conservation law to the fission process 
was highly precise was due to the above features.

We also note that, in fact, the quantity $\sum y_A m(A,Z_A)$ is independent of the incident-neutron energy until the yields 
$y_A$ change significantly near the humps of the mass distributions. The calculations reveal that, in $^{235}$U and 
$^{235}$Pu fission induced by neutrons of the fission spectrum, the deviation of $\sum y_A m(A,Z_A)$ from the values 
presented in Table 1 does not exceed 0.1 MeV.

The values of $E_{tot}$ were obtained without taking into account ternary fission. Ternary fission accompanied by the emission 
of a long-range alpha particle occurs approximately in one of 500 cases; other types of ternary fission are much less probable. 
According to estimates, the change in $E_{tot}$ upon taking into account ternary fission does not exceed 0.02\%.

In calculating $E_{tot}$, we disregarded the alpha decays of $^{144}$Nd, $^{147}$Sm and $^{149}$Sm nuclei, which are 
formed upon the completion of beta-decay processes. The total yield of these alpha-particle emitters is about 10\%; however, 
their half-lives exceed $10^{11}$ yr, so that they make no significant contribution to the energy release.

\subsection{$Effective\; Energy\; E_{eff}$}

In this subsection, we describe schematically a procedure for calculating the energies $\langle{E_\nu}\rangle$ carried away by 
antineutrinos and the corrections $\Delta E_{\beta \gamma}$ and present the results of these calculations, along with the values 
found for the effective energies $E_{eff}$ according to relation (3.1). \medskip

{\bf 1.}$\,$ Along with electron antineutrinos ($\bar{\nu_e}$) emitted by fission fragments, a considerable number of electron 
antineutrinos are generated in a reactor that are emitted in the beta decay of nuclei produced upon the activation of the materials 
occurring in the reactor by neutrons. In calculating the energy $\langle{E_\nu}\rangle$, we take into account only those reactor 
antineutrinos that are emitted by fission fragment not perturbed by the interaction with neutrons rather than all of them.

The $\bar{\nu_e}$ spectrum decreases fast with increasing energy $E_\nu$, virtually vanishing at $E_\nu \approx 10$ MeV. 
In this spectrum, the hard section $E_\nu \ge 2$ MeV contains about 60\% of the energy $\langle E_\nu \rangle$ that is carried 
away by antineutrinos.

In the case of $^{235}$U, $^{239}$Pu, and $^{241}$Pu, the $\bar{\nu_e}$ spectra necessary for calculating 
$\langle{E_\nu}\rangle$ were determined in the following way:

\begin{itemize}
\item For the region of energies above 1.8 MeV, use was made of the spectra found in the Laue$-$Langevin Institute (ILL) by 
reconstructing the measured spectra of beta electrons emitted by fission fragments [11], small corrections of about 2.5\% that 
correspond to the contributions of long-lived beta emitters [12] and which were disregarded in [11] being introduced in
these spectra.
\item The $\bar{\nu_e}$ spectra that we calculated for the energy range 0$-$3 MeV were smoothly matched in the segment 
between 2 and 2.5 MeV with the corrected ILL spectra. As a result, the calculated values changed by 2 to 3\%.
\end{itemize}

In the case of $^{238}$U, the energy $\langle{E_\nu}\rangle$ was found on the basis of the $\bar{\nu_e}$ spectrum calculated 
in the present study.

In the region $E_\nu < 2$ MeV, it is not easy to estimate the error in the energy carried away by electron antineutrinos. The 
database used in the relevant calculation includes information about 571 fission fragments. For them, the overwhelming majority 
of decay diagrams is well known. The error in determining this part of $\langle{E_\nu}\rangle$ is likely to be within 4\%.

We recall that, in fission, nuclei emit about 6 $\bar{\nu_e}$ of mean energy approximately equal to 1.5 MeV. For the fissile 
isotopes in question, the $\langle{E_\nu}\rangle$ values (in MeV/fiss.) found in the way outlined above are

$$
^{235}{\rm U} : 9.07 \pm 0.32 \qquad ^{238}{\rm U} : 11.00 \pm 0.80 \qquad \qquad \qquad \qquad \quad \;\;  
$$

\begin{equation}
^{239}{\rm Pu} : 7.22 \pm 0.27 \qquad ^{241}{\rm Pu} : 8.71 \pm 0.30
\end{equation}

We note that the errors in our knowledge of the outgoing-neutrino energies $\langle{E_\nu}\rangle$ are much greater
than the errors in determining the energies $E_{tot}$.

Part of the energy carried away by antineutrinos of energy $E_\nu \ge 1.8$ MeV can be directly compared with data obtained in 
an experiment at the reactor of the Rovno atomic power plant [13]. In that experiment, the positron spectrum was measured
in the inverse beta-decay reaction $\bar{\nu_e}+p \rightarrow n + e^+$¯ and the $\bar{\nu_e}$ spectrum was reconstructed 
in the energy region $E_\nu >$ 1.8 MeV. The value found with the aid of this spectrum for the energy that is carried away is
in satisfactory agreement with that which was calculated in the present study; that is,

\begin{equation}
X_{Rovno/calc} = 4.679/4.815 = 0.972.
\end{equation}

{\bf 2.}$\:$We recall that the energy $E_{\beta\gamma}$ released upon the complete beta decay of a pair of fission fragments is
contained in the total fission energy $E_{tot}$. The correction $\Delta E_{\beta\gamma}(t)$ takes into account the fact that, 
at the instant of observation $t$, the decay processes have not yet been completed,

\begin{equation}
\Delta E_{\beta\gamma}(t) = E_{\beta\gamma}(\infty) - E_{\beta\gamma}(t) = \int_t^\infty dt' f_{\beta\gamma}(t'),
\end{equation}
where $E_{\beta\gamma}(t)$ is the energy released per second at the instant $t$ reckoned from the beginning of the fission
process proceeding at a rate of 1 fiss./s and $f_{\beta\gamma}(t)$ is the energy released per unit time after a lapse of time $t$ 
from a single fission event [compare with the analogous expressions in (4) for $E_{tot}$].

The energy $\Delta E_{\beta\gamma}(t)$ of fission fragments that did not decay first decreases fast with increasing duration
of the irradiation of the fuel used; this decrease gradually becomes slower, with the result that, at irradiation times of about 1.5 yr, 
$\Delta E_{\beta\gamma}(t)$ virtually reaches a plateau (see Fig. 3). The formation of this plateau is associated with fragments 
whose lifetime exceeds 30 yr. Presented immediately below are the values of $\Delta E_{\beta\gamma}(t)$ (in MeV/fiss.) at 
the fuel-irradiation time corresponding to the midpoint of the standard operating period of a PWR reactor:

$$
^{235}{\rm U} : 0.35 \pm 0.02 \qquad \;\; ^{238}{\rm U} : 0.33 \pm 0.03 \qquad \qquad \qquad \qquad \qquad  
$$

\begin{equation}
^{239}{\rm Pu} : 0.30 \pm 0.02 \qquad ^{241}{\rm Pu} : 0.29 \pm 0.03.
\end{equation}

It is useful to have an analytic expression for the energy $\Delta E_{\beta\gamma}(t)$. Over a wide interval of the times $t$,
the expression

$$
^{fit}\Delta E_{\beta\gamma}(t) = E_0\, exp(-\lambda_0 t^\alpha) + \varepsilon, \qquad \qquad \qquad \qquad \qquad  
\qquad \qquad   
$$

\begin{equation}
0.5 < t < 500\; days
\end{equation}
at the $E_0$, $\lambda_0$, $\alpha$, and $\varepsilon$ values given in Table 2 agree with the results of the precise calculation 
to within 2 \%.

\begin{table}[htb]
\caption{Parameters of the functions $^{fit}\Delta E_{\beta\gamma}(t)$}
\label{table}
\vspace{5pt}
\hspace{60pt}
\begin{tabular}{c|c|c|c|c|}
\hline
Fissile nucleus & $E_0,$MeV  & $\lambda_0$ & $\alpha$ & $\varepsilon,$ MeV\\
\hline
$^{235}$U & 8.80 & 2.15 & 0.108 &\ 0.185\\
$^{238}$U & 9.20 & 2.22 & 0.106 &\ 0.165\\
$^{239}$Pu & 8.50 & 2.18 & 0.109 &\ 0.155\\
$^{241}$Pu & 8.20 & 2.16 & 0.105 &\ 0.135\\
\hline
\end{tabular}\\[2pt]
\end{table}

The first term in (15) describes an exponential decay with a decay probability decreasing with time, while the second term 
corresponds to the plateau. \medskip

{\bf 3.} To conclude this section, we present the values of the effective fission energy $E_{eff}$ (in MeV/fiss.) that
correspond to the midpoint of the reactor operating period: 

$$
^{235}{\rm U} : 193.37 \pm 0.33 \qquad ^{238}{\rm U} : 194.60 \pm 0.81 \qquad \qquad \qquad \qquad   
$$

\begin{equation}
^{239}{\rm Pu} : 199.80 \pm 0.28 \qquad ^{241}{\rm Pu} : 202.04 \pm 0.32.
\end{equation}

\section {TOTAL THERMAL ENERGY $E_f$}

In this section, we present the results obtained by calculating the energy $E_{nc}$ (in MeV/fiss.) absorbed in a reactor upon 
the capture of neutrons not involved in the chain reaction, determine the total thermal energy $E_f$, and consider its variation 
within the reactor operating period.

{\bf 1}. Of the total number $n_f$ of neutrons emitted in a fission event, only one contributes to the chain reaction. The 
remaining neutrons are absorbed almost completely in the reactor core, reflector, and vessel. The probabilities of the 
absorption of these neutrons by various substances and the energies $E_{nck}$ released in the capture of one neutron in 
those substances are quoted in Table 3.

\begin{table}[htb]
\caption{ Balance of the absorption of neutrons not involved in the chain reaction and of the thermal energy  $E_{nck}$ 
released upon the absorption of a single neutron in a given material (midpoint of the operation period)}
\label{table}
\vspace{10pt}
\begin{tabular}{c|c|c||c|c|c}
\hline
 & Capture & $E_{nck}$, & & Capture & $E_{nck}$, \\
 Material & probability  & $ \frac{MeV}{neutron}$ & Material & probability & $\frac{MeV}{neutron}$ \\
 & $\eta_k, $\% & & & $\eta_k, $\% & \\
 \hline
$^{235}$U & 11.6 & 6.54 & $^{149}$Sm & 0.8 & 7.99 \\
$^{238}$U & 38.4 & 5.72 & Other fragments & 6.8 & 7.88 \\
$^{239}$Pu & 10.5 & 6.53 & Zirconium & 7.0 & 8.11 \\
$^{240}$Pu & 6.1 & 5.24 & $^{10}$B & 5.6 & 2.79 \\
$^{241}$Pu& 3.6 & 6.31 & Water & 4.4 & 2.22 \\
$^{135}$Xe& 3.4 & 7.49 & Other materials & 1.8 & 5.67 \\
\hline
\end{tabular}\\[2pt]
\end{table}

From those data, it can be seen that more than 80\% of $(n_f-1)$ neutrons are absorbed in the fuel and in the accumulated 
fission fragments. In all cases, with the exception of that of $^{10}$B, the neutrons are absorbed via $(n,\gamma)$ reactions. 
The energies $E_{nck}$ include the energy of photons emitted in radiative neutron capture and, if beta-radioactive nuclei are 
formed, the energy of beta electrons and photons originating from the subsequent transformations of these nuclei.

The mean energy absorbed in the reactor in the capture of one neutron, $E_{n1} = \sum \eta_k E_{nck}$, and calculated on 
the basis of the data given in Table 3 is $E_{n1} = 5.97 \pm 0.15$ MeV/neutron, its increase within the period from 1 day to 
the end of the operating period being 0.55 MeV/neutron.

At \ the \ midpoint \ of \ the \ reactor \ operating \ period, \ the \ energies $E_{nci} = E_{n1}\cdot (n_{fi}-1)$ (in MeV/fiss.) entering into the 
total thermal energy of the fission of uranium and plutonium isotopes are:

$$
^{235}{\rm U} : 8.55 \pm 0.22 \qquad \quad ^{238}{\rm U} : 10.92 \pm 0.28 \qquad \qquad \qquad \qquad \quad   
$$

\begin{equation}
^{239}{\rm Pu} : 11.19 \pm 0.28 \qquad ^{241}{\rm Pu} : 11.56 \pm 0.29.
\end{equation}

{\bf 2}. We now present the values obtained for the total thermal energies $E_{fi}$ of fissile isotopes by summing the 
components found above, see Table 4.

\begin{table}[htb]
\caption{\ Thermal \ fission \ energies \ $E_{fi}$, \ at \ the \ midpoint \ of \ the \ reactor \ operating period}
\label{table}
\vspace{5pt}
\hspace{55pt}
\begin{tabular}{c|c}
\hline
Isotope & $E_{fi}$, MeV/fission \\
\hline
$^{235}$U & 201.92 $\pm$ 0.46 \\
$^{238}$U & 205.52 $\pm$ 0.96 \\
$^{239}$Pu & 209.99 $\pm$ 0.60 \\
$^{241}$Pu & 213.60 $\pm$ 0.65 \\
\hline
\end{tabular}
\end{table}

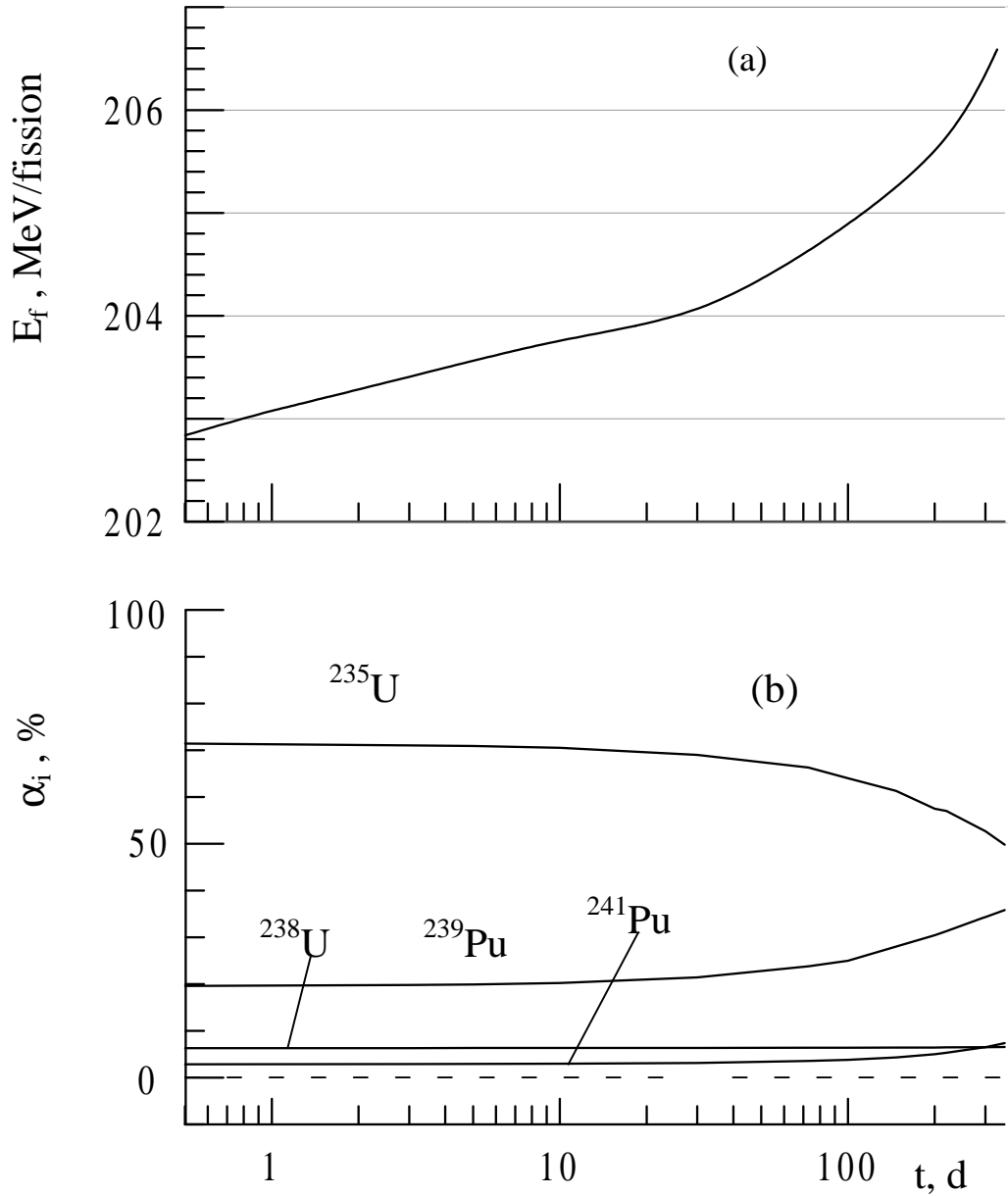
The total thermal energy $E_f = \sum \alpha_i E_{fi}$ and the contributions of fissile isotopes to the total number of fission 
events within the operating period of a PWR reactor are given in Fig. 4 versus the time of reactor operation. At the midpoint of 
the operating period, we have $E_f = 205.3$ MeV/fission. The errors in the values $E_f$ are estimated at 0.6 MeV, which 
corresponds to about 0.3\%. They include both the errors in $E_{tot}$, $\langle E_{\nu}\rangle$, $\Delta E_{\beta \gamma}$,
and $E_{nc}$ and the errors in $\alpha_i$. It is assumed that the latter are 5\% (relative errors). The increase in $E_f$ over 
the segment from 0.5 d after the start-up to the end of the operating period is 3.75 MeV. This increase is caused by three 
reasons: the growth of the energy $E_{nc}$ released in neutron capture, a decrease in the fraction of $^{235}$U and an 
increase in the contributions of $^{239}$Pu and $^{241}$Pu in the process of reactor operation, and the "start-up effect" 
that is associated with the growth of the beta- and gamma-radiation energy and which is the most sizable within the first week 
after the start-up (see Fig. 4).

\section {CONCLUDING COMMENTS}

The energy $E_f$, which relates the number of fission events occurring in a reactor \ to \ its \ thermal \ power, \ has \ been \ 
calculated \ with \ an \ error \ of $\delta E_f/E_f \approx 3 \times 10^{-3}$. The high precision of the calculation of this energy 
has been achieved owing to the possibility of finding its main component $E_{tot}$ with a relative error as small as about 
$5\cdot 10^{-4}$. The three other components, $\langle E_{\nu} \rangle$, $\Delta E_{\beta \gamma}$, and $E_{nc}$, 
have been computed to a poorer precision, but they are relatively small, not exceeding 5\% of $E_f$ .

The energy $E_f$ increases throughout the operating period. At a constant thermal power, the number of fission events in the 
reactor decreases from the beginning to the end of the operating period.

We note that all of the components appearing in expression (3) for the energy $E_f$, with the exception of $E_{nc}$, are 
characteristics of the fission of the nuclei being considered, so that their calculation is based on nuclear-physics data $-$ in 
particular, data associated with the physics of fission.

The special features of a reactor manifest themselves in the following:
\begin{itemize}
\item Use is made of the chain-reacting condition, which implies that one of the fission neutrons from the preceding generation 
induces one new fission process in the next generation.
\item Numerical data on the fission branching fractions $\alpha_i$ and on their time dependence are employed.
\item The term $E_{nc}$ is calculated with the aid of data on the balance of neutron absorption in a reactor.
\end{itemize}

As a typical example, we have presented results (see Fig. 4) concerning a standard operating period of PWR reactors, which 
are widely used in Europe, the United States of America, and Japan. However, an actual operating period of a PWR reactor may
differ from a standard one significantly. There also exist other high-power reactors at which neutrino investigations are being 
presently performed or are planned. These reactors differ from their PWR counterparts by the duration of the operating period, 
the enrichment of the nuclear fuel used, and some other special features. In all such cases, the method developed in the present 
study and the results obtained here can be used in neutrino investigations to perform a quantitative analysis of the relationship 
between the level of power and the rate of the chain reaction in the reactor core.

For the first time, thermal fission energies were calculated more than 30 years ago [14]. Later on, a new calculation was 
performed [15] in connection with neutrino investigations at the Rovno atomic power plant. In the present study, we have 
employed the most recent data concerning the issue being considered and, for the first time, have traced the dynamics
of thermal fission energy throughout the reactor operating period.

\section*{Acknowledgments}

We are grateful to M.S. Yudkevich, V.D. Sidorenko, and S.N. Bolshagov for consultations on problems in the physics 
of nuclear reactors.
 
This work is supported by the Russian Foundation of Basic Research (project no. 03-02-16055) and was also funded with 
a grant in support of leading scientific schools.


\begin{thebibliography}{99}
\bibitem {Koz}Yu. Kozlov et al., Nucl. Phys. B (Proc. Suppl. ) {\bf 87}, 514 (2000).
\bibitem {Be} A.G. Beda, E.V. Demidova, A.S. Starostin, and M.B. Voloshin, Yad. Fiz. {\bf 61}, 72 (1998) [Phys. At.
Nucl. {\bf 61}, 66 (1998)].
\bibitem {Ap}M. Appolonio et al. (CHOOZ Collab., Phys. Lett. B {\bf 420}, 397 (1998); {\bf 466}, 415 (1999).
\bibitem {Kam} KamLAND Collab., Phys. Rev. Lett. {\bf 90}, 021802 (2003).
\bibitem {Wo}H.T. Wong and J. Li  arXiv: hep-ex/0201001.
\bibitem {Scho}S. Schoenert, T. Lassere, and L. Oberauer, hep-ex/0203013; Astropart. Phys., {\bf 18}, 565 (2003).
\bibitem {De}A.V. Derbin, Fiz.@ Elem. Chastits At Yadra {\bf 32}, 739 (2001); \\ L.A. Mikaelyan, Yad. Fiz. {\bf 65}, 
1206 (2002) [Phys. At. Nucl. {\bf 65}, 1173 (2002)].
\bibitem {Au} G. Audi and A.H. Wapstra, Nucl. Phys. A {\bf 595}, 409 (1995).
\bibitem {En} T.R. England and B.F. Rider, LA-UR-94 3106, ENDF-349, LANL (Los Alamoos, 1994).
\bibitem {Ab} L.P. Abagyan et al., Vopr. At. Nauki Tekh., Ser.: Fiz. At. Reaktorov {\bf 3}, 50 (2001).
\bibitem {Sch} K. Schreckenbach et al., Phys. Lett. B {\bf 160}, 325 (1985); A. Hahn et al., Phys. Lett. B {\bf 218}, 
385 (1989).
\bibitem {Ko} V.I. Kopeikin, L.A. Mikaelyan, and V.V. Sinev, Yad. Fiz. {\bf 64}, 914 (2001) [Phys. At. Nucl. {\bf 64}, 
849 (2001)].
\bibitem {Kop} V.I. Kopeikin, L.A. Mikaelyan, and V.V. Sinev, Yad. Fiz. {\bf 60}, 230 (1997) [Phys. At. Nucl. {\bf 60}, 
172 (1997)].
\bibitem {Ja} M.F. James, J. Nucl. Energy {\bf 23}, 517 (1969).
\bibitem {Kopei} V.I. Kopeikin, Preprint IAE-4305/2 (Kurchatov \ Institute \ of \ Atomic \ Energy, Moscow, 1986).
\end{thebibliography}
\end{document}